\documentstyle[aps,prl,epsf,amsfonts]{revtex}
\newcommand{\eqb}{\begin{equation}}
\newcommand{\eqe}{\end{equation}}
\newcommand{\dmb}{\begin{displaymath}}
\newcommand{\dme}{\end{displaymath}}
\newcommand{\pad}{\partial}
\newcommand{\ep}{\varepsilon}
\newcommand{\eab}{\begin{eqnarray}}
\newcommand{\eae}{\end{eqnarray}}
\newcommand{\ra}{\right\rangle}
\newcommand{\la}{\left\langle}
\newcommand{\e}{\mbox{e}}
\newcommand{\be}{\begin{equation}}
\newcommand{\ee}{\end{equation}}

\draft

\begin{document}
\title{Vacuum structure of a modified MIT Bag}
\author{R.\ Hofmann and T.\ Gutsche}
\address{Institute of Theoretical Physics,
University of Tuebingen,
Auf der Morgenstelle 14,
72076~Tuebingen,
Germany}
\author{M.\ Schumann and R.D.\ Viollier}
\address{Institute of Theoretical Physics and Astrophysics, Department
of Physics,
University of Cape Town,
Rondebosch 7701,
South Africa}
  
\maketitle

\begin{abstract}

An alternative to introducing and subsequently 
renormalizing classical parameters in the expression for
 the vacuum energy of the MIT bag for quarks 
 is proposed in the massless case by 
 appealing to the QCD trace anomaly and scale separation due to 
 asymptotic freedom. The explicit inclusion of gluons 
 implies an unrealistically low separation scale. 
 
\end{abstract} 


The vacuum energies of spatially confined quantum fields 
have been of great interest since the early days of quantum field theory \cite{Cas,Plu}. 
Shortly after the advent of the non-Abelian gauge 
theory of strong interactions \cite{Gell-Mann,Wilczek,Politzer}, 
the  bag models of hadrons \cite{Chodos74a,Lee,SLAC} 
required estimates for the contribution of the spherically constrained 
vacuum to the total energy of a hadron. In essence,  
two lines of approaches have been pursued in the past. 
The canonical vacuum energy was parametrized by means of a
dimensionless quantity $Z_0$ to be fitted 
to experiment \cite{DeGrand}. While disregarding the quadratic boundary 
condition of the original MIT bag model, 
a relation between the bag radius $R$ and 
the bag constant $B$ was established by demanding stability of the 
calculated hadron mass under variations of $R$ 
\cite{Thomas}. However, the quadratic boundary condition 
of the fermionic MIT bag model, $B_q=\left.-\frac{1}{2}\pad_r\ (\bar{\psi}\psi)\right|_{r=R}$,
was introduced to restore the broken 
four-momentum conservation of the bag \cite{Chodos74a}, 
and thus it should be taken seriously. 
For a meaningful definition of the bag constant $B_q$ according to 
the quadratic boundary condition, 
the vacuum expectation value of this 
operator equation must be taken \cite{Milton}.   

There has been a great effort to {\em compute} the Casimir 
effect of the 
MIT bag model\cite{Milton,Milton1,Kirsten,Kirsten1,Bordag,Deutsch}. 
The vacuum expectation values of global quantities must be regularized. 
Several procedures, 
adapted to either global or local approaches, were applied. Global techniques 
regularize the sum over mode 
energies by analytical continuation (zeta-function
method) \cite{Kirsten,Kirsten1,Nes}, while local approaches compute finite 
densities based on two-point functions. The space-integral of 
these densities is regularized by volume or temporal cutoffs \cite{Bender,Plu}. 
However, different regularization schemes yield different 
answers which is not acceptable. 
Various solutions have been suggested \cite{Milton,Kirsten,Kirsten1,Bordag}. 
For instance, the vacuum energy has been separated into a classical and a quantum part. 
The classical contribution was parametrized by phenomenological quantities 
to absorb divergences due to the quantum part 
by appropriate renormalizations \cite{Kirsten,Bordag}. 
This procedure relies on direct 
experimental information which is unsatisfactory. 
Interesting results were obtained in the massive case 
\cite{Kirsten,Kirsten1,Kirsten2}. By imposing the condition 
that the vacuum of a very massive field should not fluctuate, a unique term 
in the canonical vacuum energy, attributed to quantum fluctuations, was isolated. 

In this paper we propose an alternative to the above procedure. 
Our approach is based on a separation between the 
perturbative and nonperturbative regimes of QCD. As suggested by Vepstas and Jackson 
in the framework of a chiral bag model \cite{Vepstas}, hard fluctuations 
should be allowed to traverse the boundary since 
these fluctuations are not subject to the low-energy confinement
mechanism. In contrast to the work of Ref.\,\cite{Vepstas}, 
we consider only the interior of the bag. 
In the simple model of the QCD vacuum, which the bag-model philosophy offers, 
we think of hard fluctuations to be noninteracting 
and unconfined when calculating nonperturbative effects, 
such as the ground state energy of the bag. 

\begin{figure}
\vspace{12cm}
\includegraphics{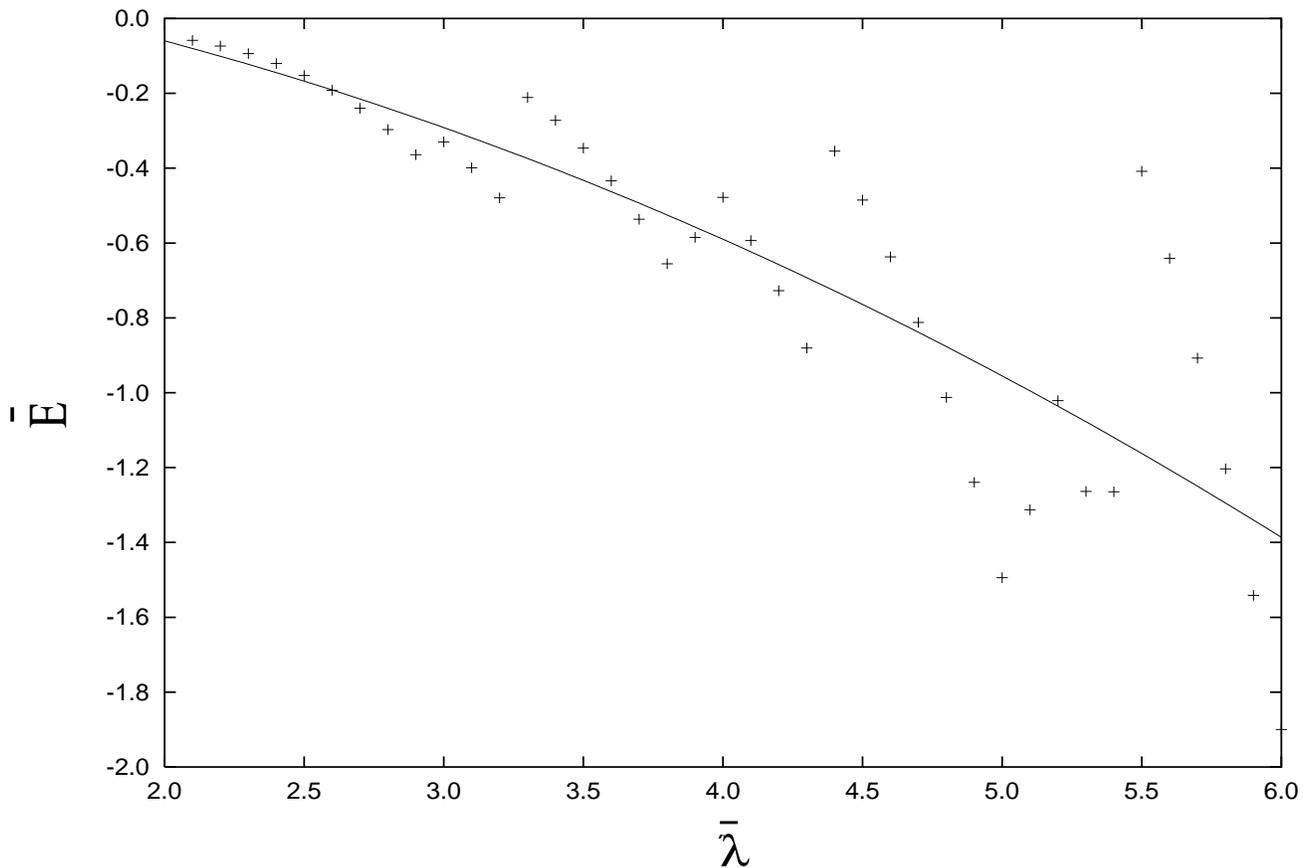}
\caption{\label{fig1}The canonical part of the one-flavor, one-color vacuum 
energy in dependence on the cutoff. Both quantities are given in units of $R^{-1}$.} 
\end{figure}

\begin{figure}
\vspace{11cm}
\includegraphics{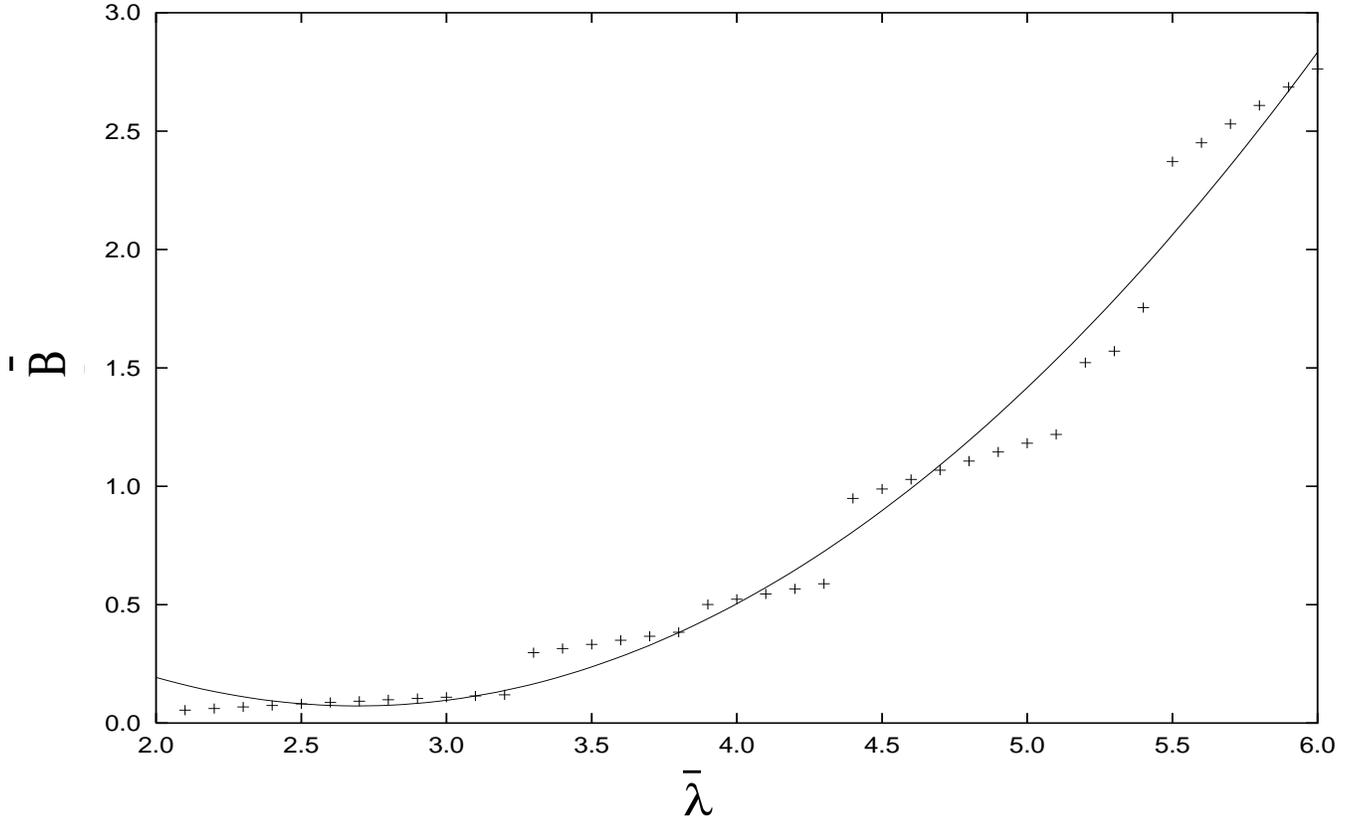}
\caption{\label{fig2}The one-flavor, one-color fermionic bag constant in dependence on 
the cutoff. The bag constant and the cutoff are given 
in units of $R^{-4}$ and $R^{-1}$, respectively.}
\end{figure}

Our numerical method to compute the regularized canonical vacuum energy 
and the bag constant of the fermionic MIT bag was explained in Ref.\,\cite{Hofmann}. 
The procedure is 
based on a mode sum representation of the cavity propagator. A Schwinger 
parametrization of the Euclidean ``momentum-squared'' denominator and a subsequent 
integration over the ``off-shell'' parameter $\omega$ are performed.
Under the condition, that 
the free-space vacuum energy vanishes, we obtain the angular 
integrated form of the canonical 
vacuum energy density $\la\tilde{\theta}^{00}\ra$ as
\eab
\label{angint}
\la\tilde{\theta}^{00}(r)\ra&\equiv&4\pi\ \la\theta^{00}(r)\ra=\frac{1}{2\ \pi^{1/2}} 
\int_{1/\lambda^2}^{\infty}dz \ \frac{1}{z^{3/2}} \nonumber\\
& &\times \Big[\sum_{\kappa}
\frac{1}{2}\sum^{n_\lambda}_{n}\frac{1}{R^3}\ {\cal N}^2_{n,\kappa}\ (2J+1)
\ \left((j_{l}(|\ep_{n,\kappa}|r))^2+
(j_{\bar{l}}(|\ep_{n,\kappa}|r))^2\right)\ 
\e^{-z \ep_{n,\kappa}^2} 
\nonumber\\ 
& &\phantom{\times \Bigg[}-\sum_{l}\ \frac{4}{\pi}\  (2l+1)
\int_0^{\lambda}dk\ k^2 (j_l(kr))^2\ \e^{-zk^2}\,\Big]\ ,\nonumber\\ 
J&=&|\kappa|-\frac{1}{2}\ ,\ \ 
l=|J|+\frac{1}{2}\ \text{sgn}\ \kappa\ ,\ \ \bar{l}=l-\mbox{sgn}\ \kappa\ .
\eae
Thereby, $j_l$ denotes the spherical Bessel function, 
and the subscripts $n$, $\kappa$, and $\mu$ stand for 
the radial quantum number, the Dirac 
quantum number, and the angular momentum projection, respectively. 
The radial quantum number $n_\lambda$ labels the mode energy closest to
$\lambda$, and  ${\cal N}^2_{n,\kappa}$ 
is a normalization constant (see Ref.\ \cite{Hofmann}). 
In Eq.\ (\ref{angint}) the integral over $k$ corresponds to 
the free-space subtraction. 
Hard fluctuations are excluded by 
distinguishing two cases: 1) hard fluctuations with 
$\omega,\ \ep_{n,\kappa}>\lambda$ or $\omega\le\lambda$, 
$\ep_{n,\kappa}>\lambda$ are omitted by 
truncation of the mode sum, and 2) hard fluctuations with $\omega>\lambda$, 
$\ep_{n,\kappa}\le\lambda$ are discarded 
by restriction of the $z$-integration. 
The canonical vacuum energy $E$ is given by 
$E=\int_{0}^{R}{\rm d}r\ r^2\ \la\tilde{\theta}^{00}(r)\ra$. 
Due to the vacuum expectation value of the 
quadratic boundary condition, the fermionic bag 
constant $B_q$ reads 
\eab
\label{Bcal}
B_q&=&-\frac{1}{4\pi^{3/2}}
\int_{1/\lambda^2}^{\infty}dz \ \frac{1}{z^{1/2}}
\sum_{\kappa}(2J+1)\sum^{n_\lambda}_{\atop{n>0}}\ \frac{1}{R^3} 
{\cal N}^2_{n,\kappa}\ \ep_{n,\kappa}^2 \e^{-z\ep_{n,\kappa}^2}\nonumber\\
& &\times \Big[\frac{j_{l}(|\ep_{n,\kappa}|R)}
{2l+1}\ \left(l\ j_{l-1}(|\ep_{n,\kappa}|R)-
(l+1)\ j_{l+1}(|\ep_{n,\kappa}|R)\right)\nonumber\\ 
& &\phantom{\times \Big[}
-\frac{j_{\bar{l}}(|\ep_{n,\kappa}|R)}
{2\bar{l}+1}\ \left(\bar{l}\,j_{\bar{l}-1}(|\ep_{n,\kappa}|R)-
(\bar{l}+1)j_{\bar{l}+1}(|\ep_{n,\kappa}|R)\right)\Big]\ .
\eae
Fig.\ \ref{fig1} shows the result of the calculation of 
$\bar{E}\equiv R\times E$ 
as a function of $\bar{\lambda}\equiv R\times \lambda$. The
discontinuous behavior is due to the fact that mode 
eigenvalues at low energies are not spaced equidistantly. 
To smooth the "nervous" behavior, we use a quadratic regression as indicated by the solid line. 
In Fig.\ \ref{fig2} the $\bar{\lambda}$ 
dependence of $\bar{B}_q\equiv R^4\times B_q$ is depicted. 
Again, a quadratic fit is used to average over discontinuities. 
Tables \ref{tab1} and \ref{tab2} contain a list of values for $3\times n_f\times B_q$, 
$-3\times n_f\times E$ under variation of $R$, where $\bar{\lambda}$ is adjusted to 
$\lambda=1.2$ GeV, $\lambda=1.6$ GeV and 
$\lambda=0.8$ GeV, $\lambda=1.0$ GeV, respectively. Thereby, 
$n_f=2$ stands for the light-flavor multiplicity, and the factor three 
is the number of colors.  

Appealing to the one-loop trace-anomaly \cite{Hatsuda} of the 
QCD energy-momentum tensor $\theta^{\mu\nu}$  
\eqb
\label{ta}
\la \theta^\mu_\mu \ra=
-\frac{1}{8}\left(11-n_f\frac{2}{3}\right)\, 
\la \frac{\alpha_s}{\pi} F_{\kappa\nu}^a F^{\kappa\nu}_a\ra\ , 
\eqe
we assume for the moment that only quark fluctuations 
contribute to the bag constant. Using the fact 
that the canonical part of $\theta^{\mu\nu}$ 
is traceless in the mixed MIT bag model, we obtain (apart from a sign) the relation  
\eqb
\label{Bta}
3\times n_f\times B_q=0.302\times \la \frac{\alpha_s}{\pi} F_{\kappa\nu}^a F^{\kappa\nu}_a\ra\ . 
\eqe
Thereby, the value of the (renormalization-scale independent) 
gluon condensate \cite{Dosch2} is 
$\la \frac{\alpha_s}{\pi} F^a_{\mu\nu}F^{\mu\nu}_a \ra=0.024\pm 0.012\ \mbox{GeV}^4$.
Comparing by means of Eq.\,(\ref{Bta}) the central value of the gluon condensate 
with the values of $3\times n_f\times B_q$ (Tables \ref{tab1}, \ref{tab2}), which are 
 stable under variation of $R$, we obtain agreement
 for $\lambda=1.0$ GeV and a bag radius $R$ of 0.6~fm. 
 Given these values of $\lambda$ and $R$, the results of  
 Table \ref{tab2} indicate that $-3\times n_f\times{E}$ is 
 close to phenomenologically obtained values: 
 In Ref.\,\cite{DeGrand} $Z_0$ parametrizes the 
Casimir energy as $-Z_0/R$. Fits to the hadron 
spectrum yield values of about $Z_0=2$ \cite{DeGrand}.  
The effect of the center-of-mass contribution to $Z_0$ 
was found to be of the order of 40\% in Refs.\,\cite{Donog,Milton2}. 
In comparison, our value of 
$-3\times n_f\times E=0.597$ GeV at $R=0.6$ fm corresponds to 
$Z_0$=1.79 with no center-of-mass contribution. 
 
How do confined gluons alter the results obtained so far? 
Analogous to the fermionic case the gluonic bag constant $8\times B_g$ is defined 
as the vacuum expectation value of the following quadratic 
boundary condition \cite{Chodos74a} 
\eqb
\label{Bgl}
B_g=-\frac{1}{4} F_{\mu\nu}F^{\mu\nu}=\frac{1}{2}(\vec{E}^2-\vec{B}^2)\ ,
\eqe
where to lowest order in the coupling the 
field strength tensor $F_{\mu\nu}$ is Abelian, and $\vec{E}$ and $\vec{B}$ denote the electric and
magnetic field strength, respectively. Appealing in the sourceless case 
to the symmetry of Maxwell's equations under the
duality transformation $\vec{E}=\vec{B}^D\ ,\ \vec{B}=-\vec{E}^D$, 
we obtain due to physical transverse polarizations (TE,TM) 
the following expression for $B_g$ in Feynman gauge
\eab
\label{Bglu}
B_{g}&=&\frac{1}{32\pi^{3/2}}\frac{1}{R^3}
\int_{1/\lambda^2}^{\infty}dz \ \frac{1}{z^{3/2}}
\sum_{n, J\ge 1}\times\nonumber\\ 
& &\left\{(2J+1)\left[({\cal N}_{n,J}^{{\tiny\mbox{TM}}})^2j_{J}^2
(\ep^{{\tiny\mbox{TM}}}_{n,J}R)\e^{-z(\ep^{{\tiny\mbox{TM}}}_{n,J})^2}-
({\cal N}_{n,J}^{{\tiny\mbox{TM}},D})^2j_{J}^2
(\ep^{{\tiny\mbox{TM}},D}_{n,J}R)\e^{-z(\ep^{{\tiny\mbox{TM}},D}_{n,J})^2}\right]\right\}+\nonumber\\ 
& &\left\{({\cal N}_{n,J}^{{\tiny\mbox{TE}}})^2\left[(J+1)j_{J-1}^2(\ep^{{\tiny\mbox{TE}}}_{n,J}R)+Jj_{J+1}^2(\ep^{{\tiny\mbox{TE}}}_{n,J}R)\right]
\e^{-z(\ep^{{\tiny\mbox{TE}}}_{n,J})^2}\right.-\nonumber\\  
& &\left.({\cal N}_{n,J}^{{\tiny\mbox{TE}},D})^2\left[(J+1)j_{J-1}^2(\ep^{{\tiny\mbox{TE}},D}_{n,J}R)+Jj_{J+1}^2(\ep^{{\tiny\mbox{TE}},D}_{n,J}R)\right]
\e^{-z(\ep^{{\tiny\mbox{TE}},D}_{n,J})^2}\right\} \ .
\eae
Thereby, the superscript $D$ indicates that the corresponding 
eigenvalue has been obtained from the linear boundary 
condition $n_\mu (F^D)^{\mu\nu}=0$ for the {\em dual} field strength, and 
${\cal N}_{n,J}^{{\tiny\mbox{TE}}}$ (${\cal N}_{n,J}^{{\tiny\mbox{TM}}}$) denotes the normalization
constant for the corresponding mode. 
For technicalities concerning Cavity QCD in 
Feynman gauge see Refs.\ \cite{Hansson,Buser}. 
In Eq.\,(\ref{Bglu}), the introduction of the Schwinger parameter $z$ 
and the subsequent truncation of $z$-integration and 
mode summation due to the subtraction of hard
fluctuations in the vacuum is analogous to the fermionic case. 
Table \ref{tab3} contains the values for $8\times B_g$ under 
variations of $R$ with $\lambda$ adjusted to $\lambda=0.8$ GeV and $\lambda=1.0$ GeV. 
For radii $R$ less than $R=0.7$ fm there is no contribution from 
the mode sum of Eq.\,(\ref{Bglu}). We find stability for $8\times B_g$ 
under a variation of $R$ at $R=0.8$ fm with 
$8\times B_g=0.0128$ GeV$^4$ for $\lambda=0.8$ GeV and with 
$8\times B_g=0.0189$ GeV$^4$ for $\lambda=1.0$ GeV. 
Appealing to the QCD trace anomaly and requiring that the total bag constant 
$B\equiv 3\times n_f\times B_q+8\times B_g$ produces the 
central value of the gluon condensate, implies $\lambda$ to be less than $\lambda=0.8$ GeV. 
As far as the properties of the lowest
light-flavor resonances are concerned, which are believed to be strongly correlated with the QCD
condensates of lowest mass-dimension, QCD sum rules \cite{SVZ} 
suggest the onset of the perturbative
regime at values of about 1.5 - 1.8 GeV$^2$ of the 
spectral continuum threshold $s_0$ \cite{Narison,Dosch,Dosch2,Dominguez}. 
This corresponds to $\lambda$=1.22 - 1.34 GeV. 
Hence, our value of $\lambda\approx 1.0$ GeV for the pure quark bag 
seems already a bit too small which might be due to 
the mode sum representation of the 
cavity propagator with implicit spatial correlations, whereas 
$s_0$ relates to plane-wave states. 
Nevertheless, it is hard to accept values of $\lambda$ 
lower than 0.8 GeV for the mixed bag.
 
In the standard fashion \cite{Cleymans,Mosel} we now estimate the critical 
temperature $T_c$ (no baryonic chemical potential $\mu$) 
of a deconfinement phase transition from the bag constant 
$3\times n_f\times B_q$. For this we take the value $n_f\times B=0.007$ GeV$^4$ 
for $\lambda=1.0$ GeV, and with  
\eqb
\label{Tc}
4B\stackrel{!}=\pi^2
T_c^4\left(\frac{8}{15}+\frac{7}{10}\right)+B
\eqe
we obtain $T_c=203.8\,\mbox{MeV}$. 
From SU(3) Yang-Mills lattice simulations one expects a 
smooth decrease of the gluon condensate for temperatures near 
$260$ MeV \cite{Miller}. Therefore, we would have 
to correct the bag radius at zero temperature towards 
higher values near the phase transition. For comparison, we determine 
the critical temperature from the phenomenological 
value $B=4.54\times 10^{-4}$ GeV$^4$ of Ref. \cite{DeGrand} as $T_c=102.8$ MeV. 
This is too low, since otherwise 
the deconfinement phase transition would have already 
been seen experimentally \cite{BM}. 

In summary, invoking asymptotic freedom and 
appealing to the QCD trace-anomaly, the linear {\em and} 
nonlinear boundary condition of the MIT bag model for quarks provide a 
reasonable agreement of the calculated canonical vacuum energy with that 
found in hadron phenomenology which makes the introduction of phenomenological 
parameters redundant. However, the explicit inclusion of gluons drives 
the separation scale down to values which are not acceptable.

\section*{Acknowledgments}

We thank K.\ Kirsten for a stimulating correspondence. 
Financial contributions from the Foundation for 
Fundamental Research (M.S.\ and R.D.V.), the Deutsches Bundesministerium fuer Bildung 
und Forschung, contract No.\ 06 Tue 887, (T.G.), 
and the Gra\-du\-ier\-ten\-kolleg 
``Struktur und Wechselwirkung von Hadronen und Kernen'' (R.H.), 
are gratefully acknowledged.

\begin{table}
\squeezetable 
\begin{tabular}{ccccccccccccccc}
$R$ [fm]  & 
\multicolumn{2}{c}{$ 0.4 $} & 
\multicolumn{2}{c}{$ 0.5 $}  & 
\multicolumn{2}{c}{$ 0.6 $}  & 
\multicolumn{2}{c}{$ 0.7 $}  & 
\multicolumn{2}{c}{$ 0.8 $}  & 
\multicolumn{2}{c}{$ 0.9 $}  & 
\multicolumn{2}{c}{$ 1.0 $} \\ 
\hline
$\bar\lambda$  & $ 2.4 $ & $ 3.2 $  & $ 3.0 $ & $ 4.1 $  & $ 3.6 $ & $ 4.9 $  & $ 4.3 $ & $ 5.7 $  & $ 4.9 $ & $ 6.5 $  & $ 5.5 $ & $ 7.3 $  & $ 6.1 $ & $ 8.1 $ \\ 
$3\times n_f\times B_q$[GeV$^4$]  & $ 0.032 $ & $ 0.053 $  & $ 0.024 $ & $ 0.079 $  & $ 0.021 $ & $ 0.089 $  & $ 0.026 $ & $ 0.088 $  & $ 0.028 $ & $ 0.082 $  & $ 0.028 $ & $ 0.075 $  & $ 0.027 $ & $ 0.068 $ \\ 
$-3\times n_f\times E$ [GeV]  & $ 0.450 $ & $ 1.060 $  & $ 0.716 $ & $ 1.439 $  & $ 0.942 $ & $ 1.779 $  & $ 1.145 $ & $ 2.095 $  & $ 1.334 $ & $ 2.398 $  & $ 1.513 $ & $ 2.690 $  & $ 1.686 $ & $ 2.976 $ \\ 
\end{tabular}
\vspace{1ex}
\caption{\label{tab1}The dependence of the fermionic bag constant and the canonical part of the
fermionic vacuum energy on the cutoff $\bar\lambda=\lambda\times R$ for two 
light-quark flavors with $R$ ranging from 0.4 fm to 1.0 fm. 
The lower and upper values of $\bar\lambda$ correspond to 
$\lambda = 1.2\ \mbox{GeV}$ and $\lambda=1.6$ GeV, respectively.}
\end{table}

\begin{table}
\squeezetable 
\begin{tabular}{ccccccccccccccc}
$R$ [fm]  & 
\multicolumn{2}{c}{$ 0.4 $} & 
\multicolumn{2}{c}{$ 0.5 $}  & 
\multicolumn{2}{c}{$ 0.6 $}  & 
\multicolumn{2}{c}{$ 0.7 $}  & 
\multicolumn{2}{c}{$ 0.8 $}  & 
\multicolumn{2}{c}{$ 0.9 $}  & 
\multicolumn{2}{c}{$ 1.0 $} \\ 
\hline
$\bar\lambda$  & $  1.6 $ & $ 2.0 $  & $ 2.0 $ & $ 2.5 $  & $2.4 $ & $ 3.0 $  & $ 2.8 $ & $ 3.5 $  & $ 3.2 $ & $ 4.1 $  & $ 3.6 $ & $ 4.6 $  & $ 4.1 $ & $ 5.1 $ \\ 
$3\times n_f\times B_q$ [GeV$^4$]  & $ 0.129 $ & $ 0.065 $  & $ 0.027 $ & $ 0.011 $  & $ 0.006$ & $ 0.007 $  & $ 0.003 $ & $ 0.010 $  & $ 0.003 $ & $ 0.012  $  & $  0.004 $ & $ 0.013 $  & $ 0.005 $ & $ 0.014 $ \\ 
$-3\times n_f\times E$ [GeV]  & $ -0.031 $ & $ 0.193 $  & $ 0.154 $ & $ 0.415 $  & $ 0.300 $ & $ 0.597  $  & $ 0.422 $ &
$ 0.755 $  & $ 0.530 $ & $ 0.900 $  & $ 0.628 $ & $ 1.034 $  & $ 0.720 $ & $ 1.162 $\\ 
\end{tabular}
\vspace{1ex}
\caption{\label{tab2} Same as in Table \ref{tab1}. 
The lower and upper values of $\bar\lambda$ correspond to 
$\lambda = 0.8\ \mbox{GeV}$ and $\lambda=1.0$ GeV, respectively.}
\end{table}

\begin{table}
\squeezetable 
\begin{tabular}{ccccccccccccccc}
$R$ [fm]  & 
\multicolumn{2}{c}{$ 0.4 $} & 
\multicolumn{2}{c}{$ 0.5 $}  & 
\multicolumn{2}{c}{$ 0.6 $}  & 
\multicolumn{2}{c}{$ 0.7 $}  & 
\multicolumn{2}{c}{$ 0.8 $}  & 
\multicolumn{2}{c}{$ 0.9 $}  & 
\multicolumn{2}{c}{$ 1.0 $} \\ 
\hline
$\bar\lambda$  & $  1.6 $ & $ 1.8 $  & $ 2.0 $ & $ 2.25 $  & $2.4 $ & $ 2.7 $  & $ 2.8 $ & $ 3.15 $  & $ 3.2 $ & $ 3.6 $  & $ 3.6 $ & $
4.05 $  & $ 4.0 $ & $ 4.5 $ \\ 
$8\times B_g$ [GeV$^4$]  & $ 0$ & $ 0 $  & $ 0 $ & $ 0 $  & $ 0$ & $ 0 $  & $ 0.0133 $ & $ 0.0205 $  & $ 0.0128 $ & $ 0.0189  $  & $ 
0.0179 $ & $ 0.0302 $  & $ 0.0191$ & $ 0.0271 $ \\ 
\end{tabular}
\vspace{1ex}
\caption{\label{tab3}The dependence of the gluonic bag constant 
on the cutoff $\bar\lambda=\lambda\times R$ with $R$ ranging from 0.4 fm to 1.0 fm. 
The lower and upper values of $\bar\lambda$ correspond to 
$\lambda = 0.8\ \mbox{GeV}$ and $\lambda=1.0$ GeV, respectively.}
\end{table}

\newpage
$\mbox{}$

\bibliographystyle{prsty}

\end{document}